\documentclass[a4paper,10pt,eqno]{article}
\usepackage{theorem}
\usepackage{latexsym,amssymb,amsfonts,amsmath}
\newtheorem{thm}{Theorem}[section]
\newtheorem{lem}[thm]{Lemma}

\theoremheaderfont{\scshape}

\newcommand{\ZM}{\mathbb{Z}}

\newcommand{\CM}{\mathbb{C}}

\newcommand{\ket}[1]{|#1\rangle}

\title{{\Large {\bf Stationary measures for the three-state Grover walk \\ with one defect in one dimension}}}
\author{
{\small Takako Endo$^{1,2}$, Hikari Kawai$^{3}$, Norio Konno$^{4}$}\\
{\scriptsize $^{1}$Department of Physics, Graduate School of Humanities and Sciences, Ochanomizu University}\\
{\scriptsize 2-1-1 Ohtsuka, Bunkyo, Tokyo, 112-0012, Japan}\\
{\scriptsize $^{2}$School of Physics and Astronomy, Monash University}\\
{\scriptsize 19 Rainforest Walk, Clayton VIC 3800, Australia}\\
{\scriptsize $^{3,4}$Department of Applied Mathematics, 
Faculty of Engineering, Yokohama National University}\\
{\scriptsize 79-5 Tokiwadai, Hodogaya, Yokohama, 240-8501, Japan}\\
}

\date{\empty }
\begin{document}
\maketitle

\par\noindent
\begin{small}
\par\noindent
{\bf Abstract}. We obtain stationary measures for the one-dimensional three-state Grover walk with one defect by solving the corresponding eigenvalue problem. We clarify a relation between stationary and limit measures of the walk.

\end{small}

\setcounter{equation}{0}
\section{Introduction \label{intro}}
The quantum walk (QW) was introduced as a quantum version of the classical random walk. The QW has attracted much attention in various fields. The review and books on QWs are Venegas-Andraca [16], Konno [9], Cantero et al. [1], Portugal [14], Manouchehri and Wang [13], for examples. 

The present paper deals with stationary measures of the discrete-time case QWs on $\ZM$, where $\ZM$ is the set of integers. The stationary measures of Markov chains have been intensively investigated, however, the corresponding study for QW has not been given sufficiently. As for stationary measures of two-state QWs, Konno et al. [11] treated QWs with one defect at the origin and showed that a stationary measure with exponential decay with respect to the position for the QW starting from infinite sites is identical to a time-averaged limit measure for the same QW starting from just the origin. Konno [10] investigated stationary measures for various cases. Endo et al. [6] got a stationary measure of the QW with one defect whose quantum coins are defined by the Hadamard matrix at $x \not= 0$ and the rotation matrix at $x = 0$. Endo and Konno [3] calculated a stationary measure of QW with one defect which was introduced and studied by W\'ojcik et al. [15]. Moreover, Endo et al. [5] and Endo et al. [2] obtained stationary measures of the two-phase QW without defect and with one defect, respectively. Konno and Takei [12] considered stationary measures of QWs and gave non-uniform stationary measures. They proved that the set of the stationary measures contains uniform measure for the QW in general. As for stationary measures of three-state QWs, Konno [10] obtained stationary measures of the three-state Grover walk. Furthermore, Wang et al. [17] investigated stationary measures of the three-state Grover walk with one defect at the origin. Endo et al. [4] got stationary measures for the three-state diagonal quantum walks without defect or with one defect. In this paper, we consider stationary measures for the three-state Grover walk with one defect introduced by Wang et al. [17] by clarifying their argument. Moreover, we find out a relation between stationary and limit measures of the walk.

The rest of the paper is organized as follows. Section \ref{def} gives the definition of our model. In Section \ref{gfm}, we present solutions of eigenvalue problem by a generating function method. In Section \ref{sm}, we obtain stationary measures and clarify a relation between stationary and limit measures for the walk. Section \ref{sum} is devoted to summary.

\section{Definition of Our Model \label{def}}
This section gives the definition of our three-state QW with one defect at the origin on $\ZM$. The discrete-time QW is a quantum version of the classical random walk with additional degree of freedom called chirality. The chirality takes values left, stay, or right, and it means the motion of the walker. At each time step, if the walker has the left chirality, it moves one step to the left, and if it has the right chirality, it moves one step to the right. If it has the stay chirality, it stays at the same position. 

Let us define
\begin{align*}
\ket{L} = 
\left[
\begin{array}{cc}
1 \\
0 \\
0 
\end{array}
\right],
\qquad
\ket{O} = 
\left[
\begin{array}{cc}
0 \\
1 \\
0 
\end{array}
\right],
\qquad
\ket{R} = 
\left[
\begin{array}{cc}
0 \\
0 \\
1  
\end{array}
\right],
\end{align*}
where $L, O$ and $R$ refer to the left, stay and right chirality states, respectively.  

The time evolution of the walk is determined by a sequence of $3 \times 3$ unitary matrices $\{ U_x : x \in \ZM \}$, where
\begin{align*}
U_x =
\left[
\begin{array}{ccc}
u_{x,11} & u_{x,12} & u_{x,13} \\
u_{x,21} & u_{x,22} & u_{x,23} \\
u_{x,31} & u_{x,32} & u_{x,33} 
\end{array}
\right],
\end{align*}
with $u_{x,jk} \in \mathbb C \> (x, \in \ZM, j,k =1,2,3)$ and $\CM$ is the set of complex numbers. To define the dynamics of our model, we divide $U_x$ into three matrices:
\begin{eqnarray*}
U_x^{L} =
\left[
\begin{array}{ccc}
u_{x,11} & u_{x,12} & u_{x,13} \\
0 & 0 & 0 \\
0 & 0 & 0 
\end{array}
\right], 
\quad
U_x^{O} =
\left[
\begin{array}{ccc}
0 & 0 & 0 \\
u_{x,21} & u_{x,22} & u_{x,23} \\
0 & 0 & 0 
\end{array}
\right], 
\quad
U_x^{R} =
\left[
\begin{array}{ccc}
0 & 0 & 0 \\
0 & 0 & 0 \\
u_{x,31} & u_{x,32} & u_{x,33} 
\end{array}
\right], 
\end{eqnarray*}
with $U_x =U_x^{L}+U_x^{O}+U_x^{R}$. The important point is that $U_x^{L}$ (resp. $U_x^{R}$) represents that the walker moves to the left (resp. right) at position $x$ at each time step. $U_x^{O}$ represents that the walker stays at position $x$. 

The model considered here is 
\begin{align*}
U_x=
\begin{cases}
\omega U_G  & (x=0),
\\
U_G & (x= \pm1, \pm2, \ldots).
\end{cases}
\end{align*}
where $\omega = e^{i \theta} \> (\theta \in [0,2\pi))$. Here $U_G$ is the Grover matrix given by
\begin{align*}
U_G
=
\frac{1}{3}
\begin{bmatrix}
-1&2&2\\
2&-1&2\\
2&2&-1
\end{bmatrix}.
\end{align*}
If $\theta = 0$, that is, $\omega =1$, then $U_x=U_G$ for any $x \in \ZM$. So this space-homogeneous model is equivalent to the usual three-state {\it Grover walk} on $\ZM$.

Let $\Psi_n$ denote the amplitude at time $n$ of the QW as follows.
\begin{align*}
\Psi_{n}
&= {}^T\![\cdots,\Psi_{n}^{L}(-1),\Psi_{n}^{O}(-1),\Psi_{n}^{R}(-1),\Psi_{n}^{L}(0),\Psi_{n}^{O}(0),\Psi_{n}^{R}(0),\Psi_{n}^{L}(1),\Psi_{n}^{O}(1),\Psi_{n}^{R}(1),\cdots ],
\\
&= {}^T\!\left[\cdots,
\begin{bmatrix}
\Psi_{n}^{L}(-1)\\
\Psi_{n}^{O}(-1)\\
\Psi_{n}^{R}(-1)
\end{bmatrix},
\begin{bmatrix}
\Psi_{n}^{L}(0)\\
\Psi_{n}^{O}(0)\\
\Psi_{n}^{R}(0)
\end{bmatrix},
\begin{bmatrix}
\Psi_{n}^{L}(1)\\
\Psi_{n}^{O}(1)\\
\Psi_{n}^{R}(1)
\end{bmatrix},
\cdots\right],
\end{align*}
where $T$ means the transposed operation. Then the time evolution of the walk is defined by 
\begin{align}
\Psi_{n+1} (x) = U_{x+1}^{L} \Psi_{n} (x+1) + U_{x}^{O} \Psi_{n} (x) + U_{x-1}^{R} \Psi_{n} (x-1).
\label{lets001}
\end{align}
Now let
\begin{align*}
U^{(s)}=
\begin{bmatrix}
\ddots&\vdots&\vdots&\vdots&\vdots&\vdots&\ldots\\
\ldots&U_{-2}^O&U_{-1}^L&O&O&O&\ldots\\
\ldots&U_{-2}^R&U_{-1}^O&U_0^L&O&O&\ldots\\
\ldots&O&U_{-1}^R&U_0^O&U_1^L&O&\ldots\\
\ldots&O&O&U_0^R&U_1^O&U_2^L&\ldots\\
\ldots&O&O&O&U_1^R&U_2^O&\ldots\\
\ldots&\vdots&\vdots&\vdots&\vdots&\vdots&\ddots
\end{bmatrix},
\;\;\;
with\;\;\;
O=\begin{bmatrix}
0&0&0\\
0&0&0\\
0&0&0
\end{bmatrix}.
\end{align*}
Then the state of the QW at time $n$ is given by
\begin{align*}
\Psi_{n}=(U^{(s)})^{n}\Psi_{0},
\end{align*} 
for any $n\geq0$. Let $\mathbb{R}_{+}=[0,\infty)$. Here we introduce a map $\phi:(\mathbb{C}^{3})^{\mathbb{Z}}\rightarrow \mathbb{R}_{+}^{\mathbb{Z}}$ such that if
\begin{align*}
\Psi= {}^T\!\left[\cdots,\begin{bmatrix}
\Psi^{L}(-1)\\
\Psi^{O}(-1)\\
\Psi^{R}(-1)\end{bmatrix},\begin{bmatrix}
\Psi^{L}(0)\\
\Psi^{O}(0)\\
\Psi^{R}(0)\end{bmatrix},\begin{bmatrix}
\Psi^{L}(1)\\
\Psi^{O}(1)\\
\Psi^{R}(1)\end{bmatrix},\cdots\right]\in(\mathbb{C}^{3})^{\mathbb{Z}},
\end{align*}
then 
\begin{align*}
\phi(\Psi) 
&= {}^T\! 
\left[ \ldots, 
|\Psi^{L}(-1)|^2 + |\Psi^{O}(-1)|^2 + |\Psi^{R}(-1)|^2, 
|\Psi^{L}(0)|^2 + |\Psi^{O}(0)|^2 + |\Psi^{R}(0)|^2, 
\right.
\\
& \qquad \qquad \left. |\Psi^{L}(1)|^2 + |\Psi^{O}(1)|^2 + |\Psi^{R}(1)|^2, \ldots 
\right] \in \mathbb{R}_{+}^{\mathbb{Z}}.
\end{align*}
That is, for any $x \in \ZM$, 
\begin{align*}
\phi(\Psi) (x) = \phi(\Psi(x))= |\Psi^{L}(x)|^2 + |\Psi^{O}(x)|^2 + |\Psi^{R}(x)|^2.
\end{align*}
Moreover we define the measure of the QW at position $x$ by
\begin{align*}
\mu(x)=\phi(\Psi(x)) \quad (x \in \ZM).
\end{align*}
Now we are ready to introduce the set of stationary measure:  
\begin{align*}
{\cal M}_{s} 
= {\cal M}_s (U)
= \left\{ \phi(\Psi_{0})\in\mathbb{R}_{+}^{\mathbb{Z}} \setminus \{ \boldsymbol{0} \} : there\;exists\;\Psi_{0}\;such\;that\;\;\phi((U^{(s)})^{n}\Psi_{0})=\phi(\Psi_{0})\;for\;any\;n\geq 0 \right\},
\end{align*}
where $\boldsymbol{0}$ is the zero vector. We call the element of ${\cal M}_{s}$ the stationary measure of the QW.

Next we consider the eigenvalue problem of the QW:
\begin{align}
U^{(s)} \Psi = \lambda \Psi \quad (\lambda \in \mathbb{C}).
\label{samui}
\end{align}
Remark that $|\lambda|=1$, since $U^{(s)}$ is unitary. We sometime write $\Psi=\Psi^{(\lambda)}$ in order to emphasize the dependence on eigenvalue $\lambda$. Then we see that $\phi (\Psi^{(\lambda)}) \in {\cal M}_s$.

Let $\mu_n (x)$ be the measure of the QW at position $x$ and at time $n$, i.e., \begin{align*}
\mu_n (x)=\phi(\Psi_n(x)) \quad (x \in \ZM).
\end{align*}
If $\lim_{n \to \infty} \mu_n (x)$ exists for any $x \in \ZM$, then we define the limit measure $\mu_{\infty} (x)$ by
\begin{align*}
\mu_{\infty} (x) = \lim_{n \to \infty} \mu_n (x) \quad (x \in \ZM).
\end{align*}

\section{Splitted Generating Function Method \label{gfm}}
In this section, we give solutions of eigenvalue problem, $U^{(s)} \Psi = \lambda \Psi$, by the splitted generating function method developed in the previous studies [11,3]. First we see that $U^{(s)} \Psi = \lambda \Psi$ is equivalent to the following relations: 
\begin{align*}
\lambda
\begin{bmatrix}
\Psi^{L}(1)\\
\Psi^{O}(1)\\
\Psi^{R}(1)
\end{bmatrix}
&=\frac{1}{3}
\begin{bmatrix}
-\Psi^{L}(2)+2\Psi^{O}(2)+2\Psi^{R}(2)\\
2\Psi^{L}(1)-\Psi^{O}(1)+2\Psi^{R}(1)\\
2\omega\alpha\ \ +\ \ 2\omega\beta\ \ -\ \ \omega\gamma
\end{bmatrix}
,
\\
\lambda
\begin{bmatrix}
\alpha\\
\beta\\
\gamma
\end{bmatrix}
&=\frac{1}{3}
\begin{bmatrix}
-\Psi^{L}(1)+2\Psi^{O}(1)+2\Psi^{R}(1)\\
2\omega\alpha\ \ -\ \ \omega\beta\ \ +\ \ 2\omega\gamma\\
2\Psi^{L}(-1)+2\Psi^{O}(-1)-\Psi^{R}(-1)
\end{bmatrix}
,
\\
\lambda
\begin{bmatrix}
\Psi^{L}(-1)\\
\Psi^{O}(-1)\\
\Psi^{R}(-1)
\end{bmatrix}
&=\frac{1}{3}
\begin{bmatrix}
-\omega\alpha\ \ +\ \ 2\omega\beta\ \ +\ \ 2\omega\gamma\\
2\Psi^{L}(-1)-\Psi^{O}(-1)+2\Psi^{R}(-1)\\
2\Psi^{L}(-2)+2\Psi^{O}(-2)-\Psi^{R}(-2)
\end{bmatrix}
,
\end{align*}
and for $x\neq-1,0,1$, 
\begin{align*}
\lambda
\begin{bmatrix}
\Psi^{L}(x)\\
\Psi^{O}(x)\\
\Psi^{R}(x)
\end{bmatrix}
=\frac{1}{3}
\begin{bmatrix}
-\Psi^{L}(x+1)+2\Psi^{O}(x+1)+2\Psi^{R}(x+1)\\
2\Psi^{L}(x)\ \ -\ \ \Psi^{O}(x)\ \ +\ \ 2\Psi^{R}(x)\\
2\Psi^{L}(x-1)+2\Psi^{O}(x-1)-\Psi^{R}(x-1)
\end{bmatrix}
,
\end{align*} 
where $\Psi^{L}(0)=\alpha, \> \Psi^{O}(0)=\beta, \> \Psi^{R}(0)=\gamma$ with $|\alpha|^2 + |\beta|^2 + |\gamma|^2 >0$.  

Here we introduce six generating functions as follows:
\begin{align*}
f^j_+ (z)=\sum_{x=1}^{\infty}\Psi^{j}(x)z^x,\ \ f^j_- (z)=\sum_{x=-1}^{-\infty}\Psi^{j}(x)z^x,\ \ (j=L,\ O,\ R).
\end{align*}
Then the following lemma was given by Wang et al. [17].
\begin{lem}
We put
\begin{align*}
A=
\begin{bmatrix}
\lambda+\dfrac{1}{3z}&-\dfrac{2}{3z}&-\dfrac{2}{3z}\\\\
-\dfrac{2}{3}&\lambda+\dfrac{1}{3}&-\dfrac{2}{3}\\\\
-\dfrac{2z}{3}&-\dfrac{2z}{3}&\lambda+\dfrac{z}{3}
\end{bmatrix},\ \ \ \ 
f_\pm(z)=
\begin{bmatrix}
f^L_{\pm}(z)\\\\
f^O_{\pm}(z)\\\\
f^R_{\pm}(z)
\end{bmatrix}
,\\\\
\text{a}_+ (z)=
\begin{bmatrix}
-\lambda \alpha\\\\
0\\\\
\dfrac{\omega z(2\alpha+2\beta-\gamma)}{3}
\end{bmatrix},\ \ \ \ 
a_- (z)=
\begin{bmatrix}
\dfrac{\omega(-\alpha+2\beta+2\gamma)}{3z}\\\\
0\\\\
-\lambda \gamma
\end{bmatrix}
,
\end{align*}
where $|\alpha|^2 + |\beta|^2 + |\gamma|^2 >0$. Then we have
\begin{align*}
Af_\pm (z)=a_\pm (z).
\end{align*}
\label{lem001}
\end{lem}
We should remark that 
\begin{align*}
\det A=\dfrac { \lambda(\lambda-1)}{3z} \bigg\{z^2+3 \bigg( \lambda+\frac{4}{3}+\frac{1}{\lambda} \bigg) z+1 \bigg\}.
\end{align*}
Then $\theta_s$ and $\theta_l (\in \mathbb{C})$ are defined by
\begin{align*}
\det A=\dfrac{\lambda(\lambda-1)}{3z}(z+\theta_s)(z+\theta_l),
\end{align*}
where $|\theta_s| \leq 1 \leq |\theta_l|$. Note that $\theta_s \theta_l=1$. Lemma \ref{lem001} gives the following lemma which was also shown by Wang et al. [17].

\begin{lem}
\begin{align*}
\Psi^{L}(x)&=
\begin{cases}
\alpha{(-\theta^{L}_{s}(+))}^x & (x\geq1),
\\\\
-\frac{(3\lambda+1)\Delta(-)\omega-6(\lambda+1)\gamma}{3(\lambda-1)}{(-\theta^{L}_{s}(-))}^{-x} & (x\leq-1),
\end{cases}
\\
\\
\Psi^{O}(x)&=
\begin{cases}
-\frac{2(\Delta(+)\omega-3\alpha)}{3(\lambda-1)}{(-\theta^{O}_{s} (+))}^x & (x\geq1), \\\\
-\frac{2(\Delta(-)\omega-3\gamma)}{3(\lambda-1)}{(-\theta^{O}_{s}(-))}^{-x} &(x\leq-1),
\end{cases}
\\
\\
\Psi^{R}(x)&=
\begin{cases}
-\frac{(3\lambda+1)\Delta(+)\omega-6(\lambda+1)\alpha}{3(\lambda-1)}{(-\theta^{R}_{s}(+))}^{x} & (x\geq1),
\\\\
\gamma{(-\theta^{R}_{s}(-))}^{-x} & (x\leq-1).
\end{cases}
\end{align*}
Here $\Delta(+)=2\alpha+2\beta-\gamma,\ \Delta(-)=-\alpha+2\beta+2\gamma$, and
\begin{align*}
\theta^{L}_{s}(+)
&=-\frac{2(\lambda+1)\Delta(+)\omega-3{{\lambda}^2} (3\lambda+1)\alpha}{3\lambda(\lambda-1)\alpha}, 
\qquad
\theta^{L}_{s}(-)=\frac{(\lambda-1)\Delta(-)\omega}{\lambda\{(3\lambda+1)\Delta(-)\omega-6(\lambda+1)\gamma\}},
\\
\theta^{O}_{s}(+)
&=\frac{\Delta(+)\omega-3 \lambda^2 \alpha}
{\lambda(\Delta(+)\omega-3\alpha)},
\qquad
\theta^{O}_{s}(-)=\frac{\Delta(-)\omega-3 \lambda^2 \gamma}
{\lambda(\Delta(-)\omega-3\gamma)},
\\
\theta^{R}_{s}(+)
&=\frac{(\lambda-1)\Delta(+)\omega}{\lambda\{(3\lambda+1)\Delta(+)\omega-6(\lambda+1)\alpha\}},
\qquad
\theta^{R}_{s}(-)=-\frac{2(\lambda+1)\Delta(-)\omega-3{{\lambda}^2} (3\lambda+1)\gamma}{3\lambda(\lambda-1)\gamma}.
\end{align*}
\label{lem002}
\end{lem}

From now on, we find out a necessary and sufficient condition for 
\begin{align*}
\theta^{L}_{s}(+)=\theta^{O}_{s}(+)=\theta^{R}_{s}(+)=\theta^{L}_{s}(-)=\theta^{O}_{s}(-)=\theta^{R}_{s}(-).
\end{align*}
First we see that $\theta^{L}_{s}(+)=\theta^{R}_{s}(-)$ and $\theta^{R}_{s}(+)=\theta^{L}_{s}(-)$ give 
\begin{align*}
(\alpha-\gamma)(\alpha+\gamma-2\beta)=0.
\end{align*}
In a similar fashion, $\theta^{O}_{s}(+)=\theta^{O}_{s}(-)$ implies
\begin{align*}
(\alpha-\gamma)(\alpha+\gamma-2\beta)(\lambda+1)(\lambda-1)=0.
\end{align*}
Moreover $\theta^{L}_{s}(+)=\theta^{O}_{s}(+)$ and $\theta^{O}_{s}(+)=\theta^{R}_{s}(+)$ give 
\begin{align*}
(\lambda+1) \bigg\{9 \alpha (\omega\Delta(+)-2\alpha){\lambda}^2-6\alpha\omega\Delta(+)\lambda-\omega\Delta(+)(2\omega\Delta(+)-9\alpha) \bigg\}=0.
\end{align*}
Similarly combining $\theta^{L}_{s}(-)=\theta^{O}_{s}(-)$ with $\theta^{O}_{s}(-)=\theta^{R}_{s}(-)$ implies
\begin{align*}
(\lambda+1) \bigg\{9 \gamma (\omega\Delta(-)-2\gamma){\lambda}^2-6\gamma\omega\Delta(-)\lambda-\omega\Delta(-)(2\omega\Delta(-)-9\gamma) \bigg\}=0.
\end{align*}
From $\theta^{L}_{s}(+)=\theta^{R}_{s}(+)$, we get 
\begin{align*}
(3\lambda+1)(\lambda+1) \bigg\{9 \alpha (\omega\Delta(+)-2\alpha){\lambda}^2-6\alpha\omega\Delta(+)\lambda-\omega\Delta(+)(2\omega\Delta(+)-9\alpha) \bigg\}=0.
\end{align*}
Furthermore, $\theta^{L}_{s}(-)=\theta^{R}_{s}(-)$ gives
\begin{align*}
(3\lambda+1)(\lambda+1) \bigg\{9 \gamma (\omega\Delta(-)-2\gamma){\lambda}^2-6\gamma\omega\Delta(-)\lambda-\omega\Delta(-)(2\omega\Delta(-)-9\gamma) \bigg\}=0.
\end{align*}

Therefore we have

\begin{lem}
A necessary and sufficient condition for 
\begin{align*}
\theta^{L}_{s}(+)=\theta^{O}_{s}(+)=\theta^{R}_{s}(+)=\theta^{L}_{s}(-)=\theta^{O}_{s}(-)=\theta^{R}_{s}(-)
\end{align*}
is that $\alpha,\ \beta,\ \gamma,$ and $\lambda (\in \CM)$ with $|\alpha|^2 + |\beta|^2 + |\gamma|^2 >0$ and $|\lambda|=1$ satisfy 
\begin{align}
&\beta
=\frac{2\omega(\alpha+\gamma)}{3\lambda+\omega},
\label{sineno001}
\\
&(\alpha-\gamma)(\alpha+\gamma-2\beta)=0,
\label{sineno002}
\\
&(\lambda+1) \bigg\{9 \alpha (\omega\Delta(+)-2\alpha){\lambda}^2-6\alpha\omega\Delta(+)\lambda-\omega\Delta(+)(2\omega\Delta(+)-9\alpha) \bigg\}=0,
\label{sineno003}
\\
&(\lambda+1) \bigg\{9 \gamma (\omega\Delta(-)-2\gamma){\lambda}^2-6\gamma\omega\Delta(-)\lambda-\omega\Delta(-)(2\omega\Delta(-)-9\gamma) \bigg\}=0.
\label{sineno004}
\end{align}
\label{lem003}
\end{lem}
We should note that a relation \eqref{sineno001} is missing in Wang et al. [17].

By Lemma \ref{lem003}, we obtain the eigenvalue $\lambda$ for $U^{(s)} \Psi = \lambda \Psi$ as follows. Here we assume that $\omega \not= 1$, that is, our QW is space-inhomogeneous.

\begin{description}
\item[(i)] $\alpha=\gamma$ case. We see that Eq. \eqref{sineno003} is equivalent to Eq. \eqref{sineno004}, since $\Delta(+)=\Delta(-)=\alpha+2\beta.$
@@\begin{description}
     \item [(a)] $\alpha \neq \beta$ case.
      If $\alpha=0$, then Eq. \eqref{sineno001} gives $\beta=0$. So we assume $\alpha\beta\gamma\neq0$. Then Eq. \eqref{sineno003} implies
\begin{align*}
&\frac{27{\alpha}^2}{(3\lambda+\omega)^2}(\lambda+1) \bigg\{3(\omega-2){\lambda}^4 +2(5\omega-3)\omega{\lambda}^3
\nonumber
\\
&\qquad \qquad \qquad \qquad +(3{\omega}^2-8\omega+3)\omega{\lambda}^2 
+2(5-3\omega){\omega}^2\lambda+3{\omega}^3(1-2\omega) \bigg\}=0.
\end{align*}
      One solution of this equation is $\lambda_1=-1$. The rest of solutions $\lambda_2,\ \lambda_3,\ \lambda_4,\ \lambda_5$ are not obtained explicitly. So we do not get stationary measures.
      
     \item[(b)] $\alpha=\beta$ case. 
     If $\alpha=0$, then Eq. \eqref{sineno001} gives $\beta=0$. So we assume $\alpha\beta\gamma\neq0$. Then Eq. \eqref{sineno001} implies $\lambda=\omega$. Eq. \eqref{sineno003} gives 
     \begin{align*}
        27\lambda{{\alpha}^2}(\lambda+1){(\lambda-1)^2}=0.
     \end{align*}
     Then we have $\lambda=-1$, since $\omega \neq 1$.
   \end{description}
\item[(ii)] $\beta=\dfrac{\alpha+\gamma}{2}$ case. 
  \begin{description}
   \item[(a)] $\beta=0$ case.
   Combining Eq. \eqref{sineno001} with $\beta=(\alpha+\gamma)/2$ gives $\alpha=-\gamma$. Then from Eq. \eqref{sineno003} and $\alpha=-\gamma$, we have 
     \begin{align*}
      9 {{\alpha}^{2}} (\lambda+1)
      \left(\lambda-\frac {\omega+\sqrt {6\omega { { (\omega-1) }^{2} } } }
      {3\omega-2} \right) \left( \lambda-\frac {\omega-\sqrt {6 \omega { { (\omega-1) }^{2} } } } {3\omega-2} \right)=0.
     \end{align*}  
Remark that Eq. \eqref{sineno003} is equivalent to Eq. \eqref{sineno004}, since $\Delta(+)=-\Delta(-)=3\alpha.$ Thus, $\alpha\neq0$ implies  
     \begin{eqnarray}
      \lambda=-1,\ \ \frac {\omega\pm\sqrt {6 \omega { { (\omega-1) }^{2} } } }
      {3\omega-2}.
     \end{eqnarray}
   \item[(b)] $\beta \neq 0$ case.
   Combining Eq. \eqref{sineno001} with $\beta=(\alpha+\gamma)/2$ gives $\lambda=\omega$. Then from Eq. \eqref{sineno003} and $\lambda=\omega$, we have 
     \begin{align*}
      27 \alpha^2 \lambda (\lambda+1) {(\lambda-1)^2} =0.
     \end{align*}
     Similarly, combining Eq. \eqref{sineno004} with $\lambda=\omega$ gives
     \begin{align*}
      27 {\gamma^2} \lambda (\lambda+1) {(\lambda-1)^2} =0.
     \end{align*}
Then $\lambda=-1$ follows from above two equations. 
@\end{description}

\end{description}
We note that $\alpha=\gamma$ and $\beta=(\alpha+\gamma)/2$ gives $\alpha=\beta=\gamma$. This case is (i-b).

\section{Stationary Measures \label{sm}}
First we obtain stationary measures for (ii-a) case with respect to the following $\lambda$: 
\begin{align*}
 \lambda(\pm)=\frac {\omega\pm\sqrt {6 \omega { { (\omega-1) }^{2} } } }
      {3\omega-2}.
\end{align*}
Then we see that for $j=L,\ O,\ R$,  
\begin{align*}
\theta^{j}_{s}(\pm) =
&
 \frac{-(3\omega+2) + 2 \sqrt{6} e^{\frac{\theta}{2}i } } {(2-3\omega) 
 (1+2 \sqrt{3(1-\cos\theta)} i)}, \quad
 \frac{-(3\omega+2) + 2 \sqrt{6} e^{\frac{\theta}{2}i } } {(2-3\omega) 
 (1-2 \sqrt{3(1-\cos\theta)} i)},
\\
&
 \frac{-(3\omega+2) - 2 \sqrt{6} e^{\frac{\theta}{2}i } } {(2-3\omega) 
 (1+2 \sqrt{3(1-\cos\theta)} i)}, \quad
 \frac{-(3\omega+2) - 2 \sqrt{6} e^{\frac{\theta}{2}i } } {(2-3\omega) 
 (1-2 \sqrt{3(1-\cos\theta)} i)}.
\end{align*}
Note that $\theta^{j}_{s}(\pm),\ (j=L,\ O,\ R\ )$ do not depend on $j$ and $\pm$, so we put $\theta_s = \theta^{j}_{s}(\pm).$ Then we get
\begin{align*}
 |\theta_s|^2=\frac{37+12\cos\theta \pm 20\sqrt{6} \cos (\theta/2)}
 {(13-12\cos\theta)^2}.
\end{align*}
We should remark that if $0\leq\theta\leq4\pi$ with $\arccos (1/3) =1.2309 \ldots \le \theta \le 4\pi - \arccos (1/3) = 11.3354 \ldots$, then 
\begin{align*}
 |\theta_s|^2=\frac{37+12\cos\theta + 20\sqrt{6} \cos (\theta/2)}
 {(13-12\cos\theta)^2} \le 1.
\end{align*}
Similarly, if $0\leq\theta\leq4\pi$ with $0 \le \theta \le 2 \pi - \arccos (1/3) = 5.0522 \ldots
$ and $2 \pi + \arccos (1/3) = 7.5141 \ldots \le \theta \le 4 \pi$, then
\begin{align*}
 |\theta_s|^2=\frac{37+12\cos\theta - 20\sqrt{6} \cos (\theta/2)}
 {(13-12\cos\theta)^2} \le 1.
\end{align*}
From Lemma \ref{lem002} for this case, we have the eigenvalues for $\lambda(\pm)$ as follows.
\begin{align*}
 \Psi^{L}(x)&=\alpha\times
  \begin{cases}
   (-\theta_s)^x & (x\geq1), \\
   \frac{(3\lambda+1)\omega -2 (\lambda+1)}{\lambda-1} (-\theta_s)^{-x} & (x\leq-1),
  \end{cases}\\\nonumber\\
 \Psi^{O}(x)&=\alpha\times
  \begin{cases}
   -2 \frac{\omega-1}{\lambda-1} (-\theta_s)^x & (x\geq1),\\
   2 \frac{\omega-1}{\lambda-1} (-\theta_s)^{-x} & (x\leq-1),
  \end{cases}\\\nonumber\\
 \Psi^{R}(x)&= \alpha\times
  \begin{cases}
   -\frac{(3\lambda+1)\omega -2 (\lambda+1)}{\lambda-1} (-\theta_s)^{x} & (x\geq1), \\
   -(-\theta_s)^{-x} & (x\leq-1).
  \end{cases}
\end{align*}
Here we note that
\begin{align*}
\frac{(3\lambda(\pm)+1)\omega-2 (\lambda(\pm)+1)}{\lambda-1}=
&
\frac{(3 e^{i\theta} -2) (\sqrt{6} e^{i \frac{\theta}{2}} +2)}{\sqrt{6} e^{i \frac{\theta}{2}} -2}, \quad 
\frac{(3 e^{i\theta} -2) (\sqrt{6} e^{i \frac{\theta}{2}} +2)}{-\sqrt{6} e^{i \frac{\theta}{2}} -2},
\\
&
\frac{(3 e^{i\theta} -2) (-\sqrt{6} e^{i \frac{\theta}{2}} +2)}{\sqrt{6} e^{i \frac{\theta}{2}} -2}, \quad 
\frac{(3 e^{i\theta} -2) (-\sqrt{6} e^{i \frac{\theta}{2}} +2)}{-\sqrt{6} e^{i \frac{\theta}{2}}-2},
\end{align*}
and
\begin{align*}
\frac{\omega-1}{\lambda(\pm)-1}=
\frac{3 e^{i\theta} -2}{-2 + \sqrt{6} e^{i \frac{\theta}{2}}}, \quad 
\frac{3 e^{i\theta} -2}{-2 - \sqrt{6} e^{i \frac{\theta}{2}}}.
\end{align*}
Therefore we obtain the following main result:
\begin{thm}
We consider the three-state Grover walk with one defect at the origin on $\ZM$. Here $\omega =e^{i\theta} ( \theta \in (0,2\pi))$ and $\alpha=-\gamma, \> \beta =0$. Then a solution of $U^{(s)} \Psi =\lambda \Psi$ with 
\begin{align*}
\lambda=\frac {\omega\pm\sqrt {6 \omega { { (\omega-1) }^{2} } } }{3\omega-2}
\end{align*}
is given by
\begin{eqnarray*}
\Psi (x)= \alpha (-\theta_s )^{|x|} \times 
\begin{cases}
\begin{bmatrix}
1\vspace{2mm}\\
-\dfrac{2(\omega-1)}{\lambda-1}\vspace{2mm}\\
-\dfrac{(3\lambda+1)\omega-2(\lambda+1)}{\lambda-1}
\end{bmatrix}
& (x\geq1),
\\
\\
\begin{bmatrix}
1\\
0\\
-1
\end{bmatrix}
& (x=0),
\\
\\
\begin{bmatrix}
\dfrac{(3\lambda+1)\omega-2(\lambda+1)}{\lambda-1}\vspace{2mm}\\
\dfrac{2(\omega-1)}{\lambda-1}\vspace{2mm}\\
-1
\end{bmatrix}
& (x\leq-1).
\end{cases}
\end{eqnarray*}
Moreover the stationary measure of the walk is 
\begin{eqnarray*}
 \mu(x)= |\alpha|^2 \times
 \begin{cases}
  |\theta_s|^{2x} \bigg\{1+(13-12\cos\theta) \bigg(\dfrac{2}{m_1} + 
  \dfrac{m_2}{m_3} \bigg) \bigg\} & (x\neq0),
\\
\\
  2 & (x=0).
 \end{cases}
\end{eqnarray*}
Here
\begin{align*}
|\theta_s|^2
=\frac{37+12\cos\theta \pm 20\sqrt{6} \cos (\theta/2)}{(13-12\cos\theta)^2},
\qquad 
m_k
=5+(-1)^{n_k} \ 2 \sqrt{6} \cos (\theta/2) \quad (k=1,2,3),
\end{align*}
with $n_k \in \{0,\ 1\}$.
\label{thm001}
\end{thm}
As a corollary, we give a relation between stationary and limit measures of the walk. If $\omega=1$, then our model becomes the usual space-homogeneous three-state Grover walk on $\ZM$. As for the limit measure for an initial state; $\Psi_0 (0)={}^T{[\tilde{\alpha},\ \tilde{\beta},\ \tilde{\gamma}]}$ and $\Psi_0 (x)={}^T{[0, 0, 0]} \> (x \neq 0)$, the following result is known (see Konno [10], for example).
\begin{eqnarray*}
\mu_{\infty} (x)=
\begin{cases}
 \{(3+\sqrt{6}) |2\tilde{\alpha}+\tilde{\beta}|^2 +(3-\sqrt{6}) | \tilde{\beta}+2 \tilde{\gamma}|^2 \\
 \hspace{30mm} -2 | \tilde{\alpha}+ \tilde{\beta}+ \tilde{\gamma}|^2 \}
 \times (49-20 \sqrt{6})^x & (x\geq1),
\\
\\
 \frac{5-2\sqrt{6}}{2} (|2\tilde{\alpha}+ \tilde{\beta}|^2+| \tilde{\beta}+2 \tilde{\gamma}|^2) & (x=0),
\\
\\
 \{(3-\sqrt{6}) |2\tilde{\alpha}+\tilde{\beta}|^2 +(3+\sqrt{6}) | \tilde{\beta}+2 \tilde{\gamma}|^2 \\
 \hspace{30mm} -2 | \tilde{\alpha}+ \tilde{\beta}+ \tilde{\gamma}|^2 \}
 \times (49-20 \sqrt{6})^{-x} & (x\leq-1).
\end{cases}\\\nonumber
\end{eqnarray*}
Combining this with the corresponding (ii-a) case ($\tilde{\alpha}=-\tilde{\gamma}, \> \tilde{\beta}=0$) gives 
\begin{eqnarray}
\mu_{\infty} (x)=
  \begin{cases}
   24{|\tilde{\alpha}|}^2 (49-20 \sqrt{6})^x & (x \not= 0),
\\
\\
   4(5-2\sqrt{6}) |\tilde{\alpha}|^2 & (x=0).
  \end{cases}
\label{enoshima001}
\end{eqnarray}
On the other hand, Theorem \ref{thm001} with $\theta=0$ and $-$ part ($n_1=1,\ n_2=0,\ n_3=1$) implies 
\begin{eqnarray}
 \mu(x)= 
  \begin{cases}
  12 {|\alpha|^{2}} (5+2\sqrt{6}) {(49-20\sqrt{6})}^{|x|} & (x\neq0), 
\\
\\
2{|\alpha|^2} & (x=0).
  \end{cases}
\label{enoshima002}
\end{eqnarray}
If we put $\alpha= \pm (2-\sqrt{6}) \tilde{\alpha}$, then a stationary measure given by Eq. \eqref{enoshima002} is equivalent to a limit measure given by Eq. \eqref{enoshima001}.

Finally, we give stationary measures for $\lambda=-1$. Remark that we see $\theta_s=-1$ for this case. 
\begin{description}
 \item[(i)] $\alpha=\gamma$ case. 
  \begin{eqnarray*}
   \Psi(x)=\begin{cases}
    \begin{bmatrix}
     \alpha\\
     \frac{\alpha}{\omega-3} (3 {\omega}^2-2\omega+3)\\
     \frac{\alpha\omega}{\omega-3} (1-3\omega)
    \end{bmatrix} & (x\geq1),
\\\\
     \begin{bmatrix}
      \alpha\\
      \frac{4\omega\alpha}{\omega-3}\\
      \alpha
     \end{bmatrix} & (x=0),
\\\\
    \begin{bmatrix}
     \frac{\alpha\omega}{\omega-3} (1-3\omega)\\
     \frac{\alpha}{\omega-3} (3 {\omega}^2-2\omega+3)\\
     \alpha
    \end{bmatrix} & (x\leq-1).
   \end{cases}
  \end{eqnarray*}
  Therefore the corresponding stationary measure is given by 
\begin{eqnarray*}
 \mu(x)= \frac{6 {|\alpha|}^2}{5-3 \cos\theta}\times
  \begin{cases}
   3 \cos^2 \theta -3\cos\theta+2 & (x\neq0),
\\
\\
   3 - 2\cos\theta & (x=0).
  \end{cases}
\end{eqnarray*}

\item[(ii)] $\beta=\frac{\alpha+\gamma}{2}$ case.
 \begin{description}
  \item[(a)] $\beta=0$ case. 
   \begin{eqnarray*}
    \Psi(x)=\begin{cases}
     \begin{bmatrix}
     \alpha\\
     \alpha(\omega-1)\\
     -\omega\alpha
     \end{bmatrix} & (x\geq1),\\\\
     \begin{bmatrix}
     \alpha\\
     0\\
     -\alpha
     \end{bmatrix} & (x=0),\\\\
     \begin{bmatrix}
     \omega\alpha\\
     -\alpha(\omega-1)\\
     -\alpha
     \end{bmatrix} & (x\leq-1).
    \end{cases}
   \end{eqnarray*}
Therefore the corresponding stationary measure is given by 
\begin{eqnarray*}
 \mu(x)=2 {|\alpha|^2}\times 
  \begin{cases}
  (2-\cos\theta) & (x\neq0), \\\\
  1 & (x=0).
  \end{cases}
\end{eqnarray*}

\item[(b)] $\beta\neq0$ case.
\begin{eqnarray*}
 \Psi(x)=\begin{cases}
  \begin{bmatrix}
   \alpha\\
   -2\alpha\\
   \alpha
  \end{bmatrix} & (x\geq1),\\\\
  \begin{bmatrix}
   \alpha\\
   \frac{\alpha+\gamma}{2}\\
   \gamma
  \end{bmatrix} & (x=0),\\\\
  \begin{bmatrix}
   \gamma\\
   -2\gamma\\
   \gamma
  \end{bmatrix} & (x\leq-1).
 \end{cases}
\end{eqnarray*} 
Therefore the corresponding stationary measure is given by 
\begin{eqnarray*}
 \mu(x)=\begin{cases}
  6{|\alpha|^2} & (x\geq1),\\\\
  \dfrac{5}{4} ({|\alpha|}^2+{|\gamma|^2})+\dfrac{1}{4}
  (\alpha \bar{\gamma}+\bar{\alpha} \gamma) & (x=0),\\\\
  6{|\gamma|^2} & (x\leq-1).
 \end{cases}
\end{eqnarray*}
\end{description}
\end{description}

\section{Summary \label{sum}}
We obtained stationary measures for the three-state Grover walk with one defect at the origin on $\ZM$ by solving the corresponding eigenvalue problem. Moreover, we found out a relation between stationary and limit measures of the walk. As a future work, it would be interesting to investigate the relation between stationary measure, (time-averaged) limit measure, and rescaled weak limit measure [7,8] for QWs in the more general setting.

\par
\
\par\noindent
{\bf Acknowledgment.} This work is partially supported by the Grant-in-Aid for Scientific Research (Challenging Exploratory Research) of Japan Society for the Promotion of Science (Grant No.15K13443). 

\par
\
\par

\begin{small}
\bibliographystyle{jplain}

\end{small}

\end{document}